\documentclass[aps,prc,reprint,floatfix,showkeys]{revtex4-2}

\usepackage{float}
\usepackage{amsmath,mathtools}
\usepackage{booktabs}
\usepackage{tabularx}
\usepackage{multirow}
\usepackage{parskip}
\usepackage{hyperref}
\usepackage{graphicx}
\hypersetup{
	colorlinks=true,
	linkcolor=blue,
	citecolor=blue,
	urlcolor=blue
}
\graphicspath{ {images/} }

\newcommand{\RN}[1]{\textup{\uppercase\expandafter{\romannumeral #1}}}

\begin{document}
    \title{Mass-dependent sequential freeze-out of hadrons}
    \author{Sayan Mitra}
    \affiliation{Indian Institute of Science Education and Research, Kolkata}
    \email{sm15ms156@iiserkol.ac.in}
    \email{masterfelu@hotmail.com}
    \begin{abstract}
        Fluctuation measurements of hadron yields at heavy-ion collisions can reproduce the phase transition parameters of QCD matter. The fluctuation results produce accurate parameters near zero baryonic chemical potential($\mu_B$) being very sensitive in that region. In this work, using the Hadron Resonance Gas Model(HRG), we determine the freeze-out temperatures of the hadrons using individual fluctuation results and find that sequential freeze-out of hadrons during the QCD phase transition near zero $\mu_B$ is dependent on the mass of produced hadronic species. Using recently published results of net-$\Lambda$ fluctuations and some other fluctuation results, we study the phase diagram at low baryonic chemical potential and center-of-mass energies ranging from 19.6 -- 200 GeV/$c^2$ and find significant, quantitative evidence to support our claim.
    \end{abstract}
    
    \maketitle
	\section{Introduction}
	The investigation of the phase structure of QCD has gained considerable importance over the last several years\cite{BraunMunzinger:2009zz}. The present consensus believes in the existence of a critical point (CP) in the temperature-baryonic chemical potential $\left(T-\mu_B\right)$ phase space where the first-order phase transition of hadronic to partonic phase for $\mu_B\neq 0 $ changes to critical crossover for $\mu_B = 0$\cite{Aoki:2006we}. However, any concrete experimental evidence is yet to be determined as the confinement of QCD matter prevents us from observing partonic matter. Nevertheless, by studying the experimental results from the hadronic side, the temperature and chemical potentials at which the phase transition occurs, known as the chemical freeze-out parameters, have been determined in several works, most notably in a work by Cleymans et al. .\cite{Cleymans:2005xv}. Using the Hadron Resonance Gas Model\cite{Cleymans:2005xv,BraunMunzinger:2003zd,Bazavov:2012jq}, the experimental yields of different particles have been used to quantify the chemical freeze-out parameters. However, the parameters determined using yields fail to be precise enough in the region of CP-phase transition.
	
	The higher-order cumulants of hadron yields are very sensitive in the critical region because they scale with higher orders of correlation length\cite{Stephanov:2008qz}. The second phase of Beam Energy Scan(BES-II) at Relativistic Heavy-Ion Collider (RHIC) at STAR, BNL, that is designed to probe regions of interest in the QCD phase diagram by systematically varying the center-of-mass energy $\left(\sqrt{s_{NN}}\right)$ of the collisions (see Ref. \cite{Bzdak:2019pkr} for a recent review), has provided a wealth of data to determine the freeze-out parameters, comprising of higher-order cumulants of hadron yields. Previously, the first and second-order fluctuations of proton and kaon number measured experimentally\cite{Adamczyk:2013dal,Adamczyk:2017wsl} was used to determine the freeze-out parameters, along with net-charge moments\cite{Adamczyk:2014fia}. In this work, we have used the recently published net-lambda fluctuations\cite{Adam:2020kzk}, as well as net-kaon and net-proton, to determine the freeze-out parameters and have found compelling evidence of sequential freeze-out\cite{Bellwied:2017uat}.
	\section{Methodology}
	The HRG model can well describe the bulk-behavior of a hadronic medium in thermal equilibrium\cite{Huovinen:2009yb,Bazavov:2012jq}. The partition function  of a system of non-interacting hadrons having temperature T, volume V, baryon chemical potential $\mu_B$, charge chemical potential $\mu_Q$ and strangeness chemical potential $\mu_S$, in HRG model, is given as:
	\begin{multline}
	    \label{eq:partitionFunction}
		\ln Z\left( T, V, \mu_B, \mu_Q, \mu_S \right) =  \sum_i (-1)^{d_i} \frac{d_i V}{(2\pi)^3} \int  d^3\vec{p} \\
		\times \ln\left[1+(-1)^{d_i} \exp\big\{-\left(E_i-\mu_i\right) /T\big\}\right]
	\end{multline}
	The sum is over all the hadrons and their resonances. $E_i$ is the relativistic energy of a particle, $\sqrt{\left|\vec{p}\right|^2+m_i^2}$, having mass $m_i$, baryon number $B_i$, charge $Q_i$, strangeness $S_i$, and degeneracy $d_i$. The chemical potential, $\mu_i$ is given by $\mu_B B_i + \mu_Q Q_i + \mu_S S_i$. We consider all the established particles from 2014 PDG compilation\cite{Agashe:2014kda}.The pressure may be calculated through:
	\begin{multline}
		\label{pressure}
			P\left( T, \mu_B, \mu_Q, \mu_S \right) = \frac{T}{V}\ln Z = \sum_i (-1)^{d_i} \frac{d_i T}{(2\pi)^3} \\
			\times \int  d^3\vec{p}\,\, \ln\left[1+(-1)^{d_i} \exp\big\{-\left(E_i-\mu_i\right) /T\big\}\right]	
	\end{multline}
	The susceptibilities of conserved charges of order $n$ is defined as:
	\begin{equation}
		\chi_n^X = \frac{\partial^n{\left(P/T^4\right)}}{\partial{\left(\mu_X/T\right)}^n}
	\end{equation}
	Where $X$ is $B$, $Q$ or $S$. For individual hadrons, the derivative is done with respect to its chemical potential. The analytical expressions for first, second and third order susceptibilities for i-th hadron is given by\cite{Fu:2013gga}:
    \begin{align}
        \chi_1^{(i)} &= \frac{1}{V T^3}\left[ \left( \frac{\partial}{\partial \mu_i} \right) \ln Z \right]_{T,V} = \frac{d_i}{(2\pi)^3} \int  d^3\vec{p}\ n_i \\
        \begin{split}
            \chi_2^{(i)} &= \frac{1}{V T^3}\left[ \left( \frac{\partial}{\partial \mu_i} \right)^2 \ln Z \right]_{T,V}\\
            &= \frac{d_i}{(2\pi)^3} \int  d^3\vec{p}\ n_i\left[1+(-1)^{d_i}n_i\right]
        \end{split} \\
        \begin{split}
            \chi_3^{(i)} &= \frac{1}{V T^3}\left[ \left( \frac{\partial}{\partial \mu_i} \right)^3 \ln Z \right]_{T,V}\\
            &= \frac{d_i}{(2\pi)^3} \int  d^3\vec{p}\ n_i \left[1-(-1)^{d_i}3 n_i + 2 n_i^2\right]
        \end{split}
    \end{align}
    where
    \begin{equation}
        n_i = \frac{1}{(-1)^{d_i}+\exp\left[(E_i-\mu_i)/T\right]}
    \end{equation}
	In order to account for experimental rapidity($y$) and transverse momentum($p_T$) cuts, the integral is modified as
	\begin{equation}
		\begin{split}
			\int  d^3\vec{p} = 2\pi \int_{y^{min}}^{y^{max}} dy \int_{p_T^{min}}^{p_T^{max}} dp_T \\
			\times p_T\sqrt{p_T^2+m_i^2}\cosh (y)
		\end{split}
	\end{equation}
	The energy, $E_i$ is written as $p_T\sqrt{ p_T^2 + m_i^2 } \cosh (y)$. \\
	Experimentally, the fluctuations of the conserved charges have been measured by the STAR collaboration at BNL\cite{Adamczyk:2013dal,Adamczyk:2017wsl,Adamczyk:2017wsl} for different center-of-mass energies and centrality. The kinematic acceptance ranges specified by them, that has been used by us for HRG calculations are given as: 
	\begin{table}[H]
		\centering	
		\begin{tabularx}{0.85\linewidth}{ *6{>{\centering\arraybackslash}X} }
			\toprule\toprule
			\multirow{2}{*}{Obs.} & \multirow{2}{*}{ref.} & \multicolumn{2}{c}{$p_T$ cuts} & \multicolumn{2}{c}{y cuts} \\
			\cmidrule(r){3-4} \cmidrule(l){5-6}
			& & min & max & min & max \\
			\midrule
			net-p & \cite{Adamczyk:2013dal} & 0.4 & 0.8 & -0.5 & 0.5 \\
			net-k & \cite{Adamczyk:2017wsl} & 0.2 & 1.6 & -0.5 & 0.5 \\ 
			net-$\Lambda$ & \cite{Adam:2020kzk} & 0.9 & 2.0 & -0.5 & 0.5 \\
			\bottomrule\bottomrule
		\end{tabularx}
	\end{table}
	\begin{table*}
		\begin{tabularx}{\linewidth}{>{\hsize=0.7\hsize\linewidth=\hsize\centering\arraybackslash}X *6{>{\hsize=1.2\hsize\linewidth=\hsize\centering\arraybackslash}X} *3{>{\hsize=0.7\hsize\linewidth=\hsize\centering\arraybackslash}X}}
			\toprule\toprule
			\multirow{2}{\linewidth}{\centering $\sqrt{s_{NN}}$ (GeV)} & \multicolumn{3}{c}{$T_{ch}$ (MeV)} & \multicolumn{3}{c}{$\mu_{B,ch}$ (MeV)} & \multicolumn{3}{c}{$\chi^2$ error $\left(\times 10^{-5}\right)$}\\
			\cmidrule(l){2-4} \cmidrule(l){5-7} \cmidrule(l){8-10}
			& \multicolumn{1}{c}{k} & \multicolumn{1}{c}{p} & \multicolumn{1}{c}{$\Lambda$} & \multicolumn{1}{c}{k} & \multicolumn{1}{c}{p} & \multicolumn{1}{c}{$\Lambda$} & \multicolumn{1}{c}{k} & \multicolumn{1}{c}{p} & \multicolumn{1}{c}{$\Lambda$} \\ 
			\midrule
			19.6 & $ 144.1 \pm 0.6 $ & $ 154.9 \pm 0.1 $ & $ 156.7 \pm 0.9 $ & $ 311.4 \pm 2.2 $ & $ 205.7 \pm 0.3 $ & $ 212 \pm 2.5 $ & 0 & 0.2 & 0 \\
			27 & $ 146.8 \pm 0.8 $ & $ 159.7 \pm 2.4 $ & $ 158.1 \pm 1.5 $ & $ 249.9 \pm 2.8 $ & $ 154.7 \pm 2 $ & $ 158.1 \pm 1.6 $ & 667.8 & 636.8 & 667.8 \\
			39 & $ 145.1 \pm 0.8 $ & $ 165 \pm 1.1 $ & $ 158.8 \pm 2.6 $ & $ 203.5 \pm 2.7 $ & $ 111.7 \pm 1.9 $ & $ 115.9 \pm 4.2 $ & 945.5 & 7.3 & 945.5 \\
			62.4 & $ 142.9 \pm 2.3 $ & $ 163.4 \pm 1.4 $ & $ 158.8 \pm 24.9 $ & $ 143.4 \pm 1.7 $ & $ 73.2 \pm 1.1 $ & $ 75.2 \pm 23.1 $ & 720.9 & 1234.6 & 720.9 \\
			200 & $ 138 \pm 67.7 $ & $ 153.6 \pm 5.2 $ & $ 153 \pm 1.2 $ & $ 62 \pm 723.3 $ & $ 25.9 \pm 0.8 $ & $ 25 \pm 0.5 $ & 279.7 & 65.9 & 279.7 \\
			\bottomrule\bottomrule
		\end{tabularx}
		\caption{Freeze-out parameters obtained by fitting different hadron fluctuations. The $\chi^2$ error is the minimum value of \eqref{eq:chi2}.}
		\label{table:freezeout}
	\end{table*}
	Besides experimental cuts, we also take into account the resonance decay in a similar manner as Ref.\cite{Nahrgang:2014fza}
	\begin{equation}
		\chi_n^{(i)} = \hat{\chi}_n^{(i)} + \sum_{j}\hat{\chi}_n^{(j)} {\left\langle n_i \right\rangle}_j^n 
	\end{equation}
	Where $\hat{\chi}_n^{(i)}$ indicates the value without decay effect. We only use the average influence of the resonance decays since it agrees best with the experimental data for first and second order susceptibilities. They are related to the cumulants of the number distribution of the particle species as:
	\begin{gather}
		M = \left\langle N \right\rangle = VT^3 \chi_1 \\
		\sigma^2 = \left\langle (\Delta N)^2 \right\rangle = VT^3 \chi_2 \\
		S\sigma^3 = \left\langle (\Delta N)^3 \right\rangle = VT^3 \chi_3 \\
		\kappa\sigma^4 = \left\langle (\Delta N)^4 \right\rangle - 3{\left\langle (\Delta N)^2 \right\rangle}^2 = VT^3 \chi_4
	\end{gather}
	Where $M$, $\sigma$, $S$ and $\kappa$ are the mean, standard deviation, skewness and kurtosis of the observed yield distributions. To cancel the unknown volume term, the ratios of susceptibilities are considered. Since the correlation between a particle and its anti-particle is zero in the HRG model, the net susceptibility can be expressed as:
	\begin{equation}
		\chi_n^{(\text{net},i)} = \chi_n^{(i)} + (-1)^n \chi_n^{(\bar{i})}
	\end{equation}
	Where $\bar{i}$ denotes the index of anti-particle of $i$-th hadron. The chemical freeze-out parameters $\{T,\mu_B,\mu_Q,\mu_S\}$ are determined by the following conditions:
	\begin{align}
	    \label{eq:fitcond1}
		\left(\frac{\sigma^2}{M}\right)^{(\text{net},i)}_{\text{exp}} &= \left(\frac{\sigma^2}{M}\right)^{(\text{net},i)}_{\text{HRG}} \equiv \frac{\chi_2^{(\text{net},i)}}{\chi_1^{(\text{net},i)}} \\
		\label{eq:fitcond2}
		\langle n_S \rangle &= 0 \\
		\label{eq:fitcond3}
		\langle n_Q \rangle &= 0.4 \langle n_B \rangle 
	\end{align} 
	We perform a $\chi^2$ minimisation to determine the fit-parameters $\{T,\,\mu_B,\mu_Q,\mu_S\}$. \eqref{eq:fitcond1} determines the primary freeze-out condition ${T,\mu_B}$, and \eqref{eq:fitcond2} and \eqref{eq:fitcond3} constrains the charge and strangeness chemical potential. The $\chi^2$ function that is minimized is defined as:
	\begin{equation}
		\label{eq:chi2}
		\begin{split}
			\chi^2 &= \chi^{\text{exp.}} + \chi^{\text{constr.}}\\
			&= \begin{multlined}
				\frac{\left(\frac{\chi_2^{(\text{net},i)}}{\chi_1^{(\text{net},i)}}-\left(\frac{\sigma^2}{M}\right)^{(\text{net},i)}_{\text{exp}}\right)^2}{\sigma_{\sigma^2/M}^2} \\
				+ \left|\frac{\left<n_S\right>}{\left<n_S\right>_+ + \left<n_S\right>_-}\right| + \left|\frac{\frac{\left< n_Q\right>}{\left<n_B\right>}-0.4}{0.4}\right|
			\end{multlined}
		\end{split}
	\end{equation}
	The $\chi^2$ function consists of two parts -- $\chi^{\text{exp.}}$, which is used to determine the $T$ and $\mu$'s for which the HRG calculations match the experimental results, and  $\chi^{\text{constr.}}$, to ensure that \eqref{eq:fitcond2} and \eqref{eq:fitcond3} are satisfied. We report the point where the $\chi^2$ is least and the $1\sigma$ errors at that point.
	
	The value $\sigma_{\sigma^2/M}^2$ has been reported experimentally. $\left<n_{B/Q/S}\right>$ denotes the net-baryon/electric charge/strangeness density of the system and $\left<n_S\right>_{+/-}$ denote the density of  positively/negatively strange particles. The functional form used for $\chi^{\text{constr.}}$ has been discussed in Ref.\cite{Wheaton:2011rw}, except that instead of using Broyden convergence method, we have included the constraints in the $\chi^2$ minimization.
	\begin{figure}
		\centering
		\includegraphics[width=1\linewidth]{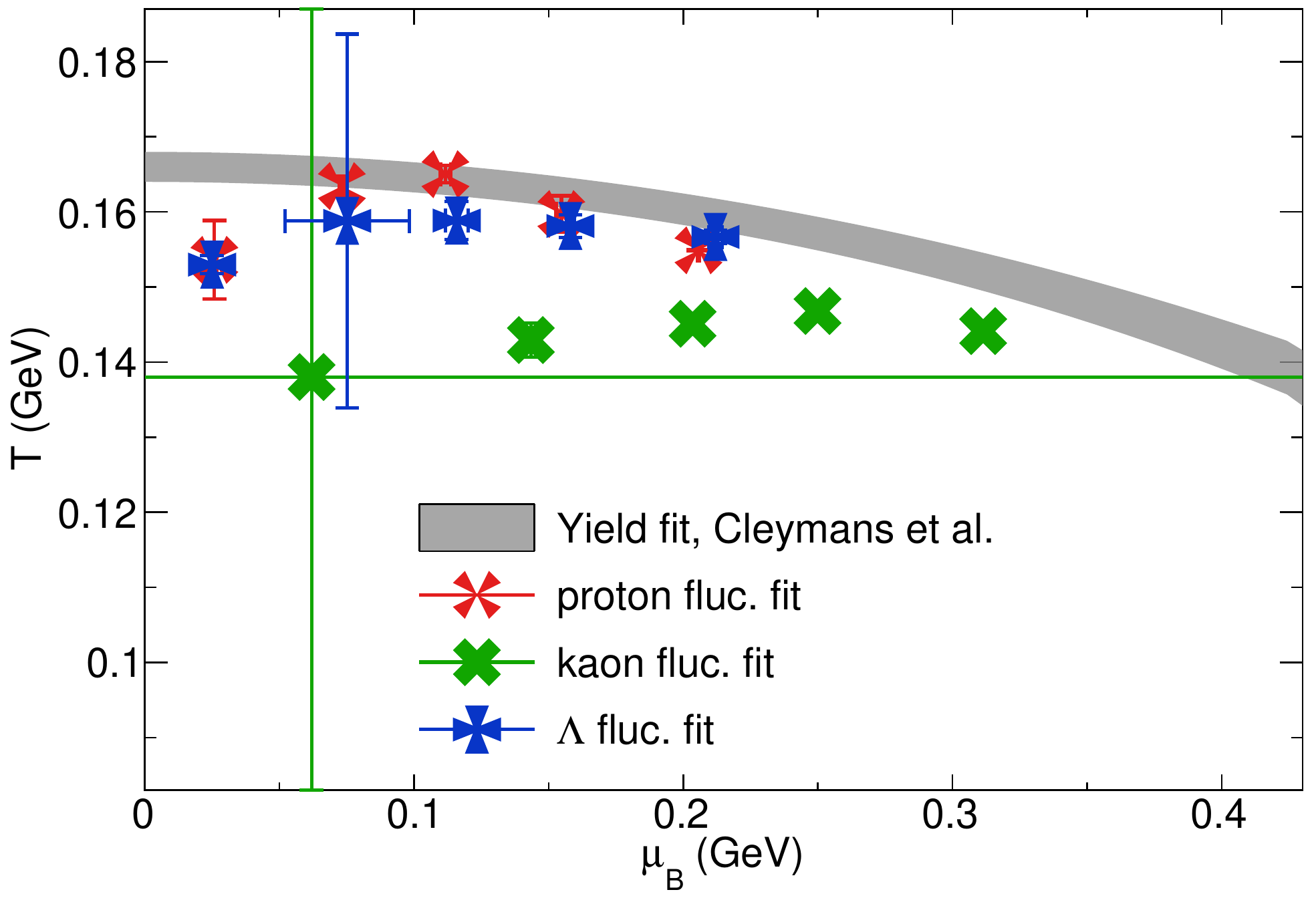}
		\caption{Freeze-out parameters obtained by fitting cumulant ratios. The large error for the lowest $\mu_B$ point for net-kaon is due to large experimental error. The yield fit is the accepted trend of freeze-out curve\cite{Cleymans:2005xv}.}
		\label{fig:freezeout}
	\end{figure}
	\begin{figure}
		\centering
		\includegraphics[width=1\linewidth]{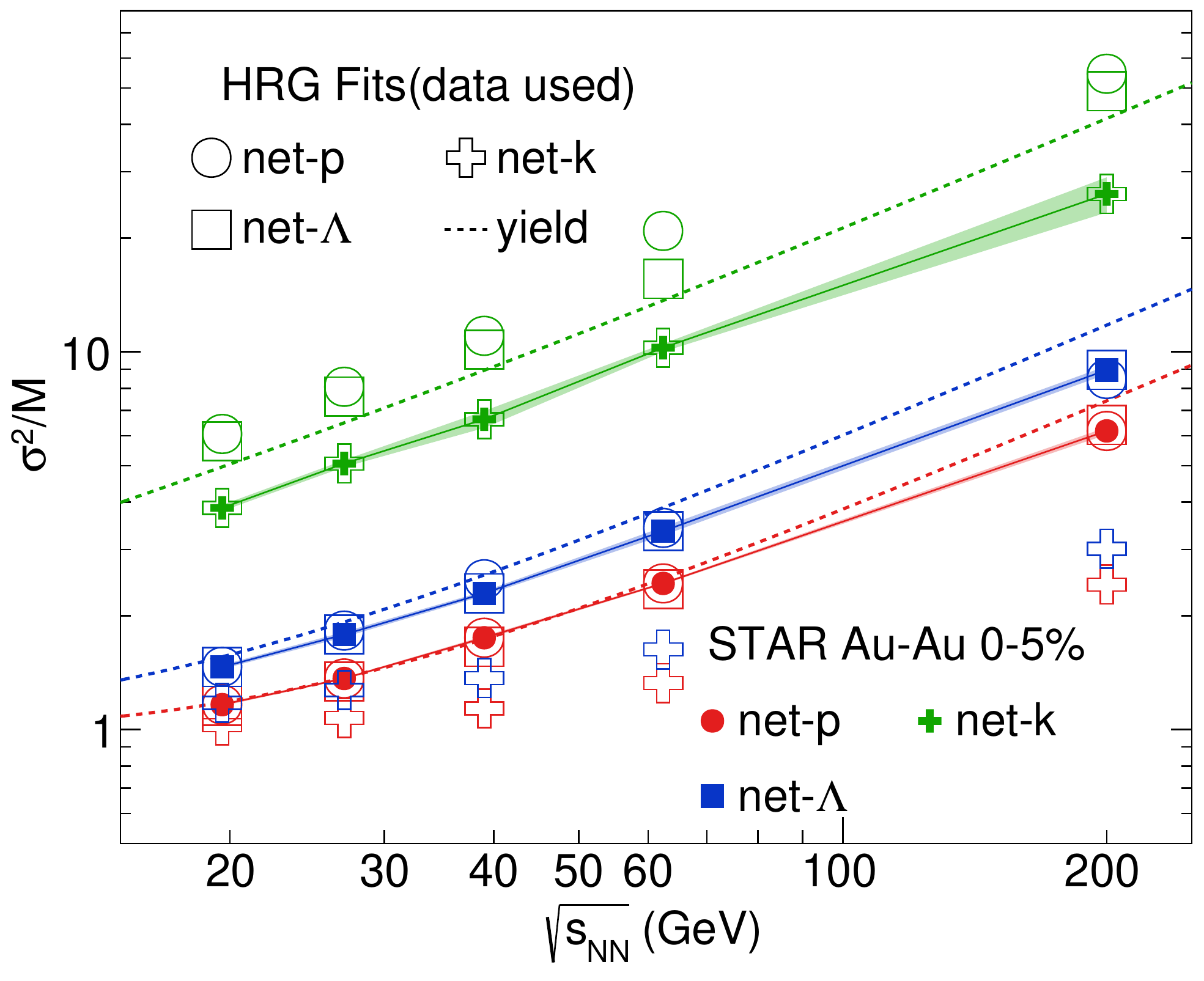}
		\caption{Plot of the fit results along with the experimental data used. The filled shapes are the experimental results and the hollow shapes are the HRG estimates using corresponding experimental results. The shaded line is the experimental error. The yield fit results are taken from the work by Cleymans et al.\cite{Cleymans:2005xv}.}
		\label{fig:FitResults}
	\end{figure}	
	\section{Results}
	We performed multiple fits using experimentally observed cumulant ratios of individual hadrons and compared the obtained freeze-out parameters in the $\{T,\mu_B\}$ plane with the accepted trend of the freeze-out curve\cite{Cleymans:2005xv} The freeze-out parameters are plotted in Fig. \ref{fig:freezeout}and tabulated in Table \ref{table:freezeout}, and the HRG estimates for the cumulant ratios using the corresponding fit results are plotted in Fig. \ref{fig:FitResults}.
	
	The $\chi^2$ error ranges from $10^{-2}$ -- $10^{-5}$. Compared to previous works which determined freeze-out parameters from yield and fluctuation observables\cite{Cleymans:2005xv,Alba:2014eba,Bluhm:2018aei}, our error is small as we have used the fluctuation results of only a single hadron. We have also observed that the $\chi^2$ error (as calculated using \eqref{eq:chi2}) is mostly governed by the strangeness constraint of $\chi^{\text{constr.}}$ \textit{i.e} the value $\left|\left<n_S\right>/\left(\left<n_S\right>_+ + \left<n_S\right>_-\right)\right|$ was maximum.
	
	We observe that for the heavier baryons(proton and lambda), the freeze-out parameters strictly follow the trend of the curve obtained by yield fit\cite{Cleymans:2005xv}. However, for kaon, which is a light, strange meson, the freeze-out temperatures are lower. We also see from Fig. \ref{fig:FitResults} that for proton and lambda, they can faithfully reproduce the experimental results of each other. However, they can neither reproduce the kaon results nor the freeze-out parameters extracted from kaon fluctuations can reproduce proton and lambda experimental observables. The generally accepted freeze-out trend\cite{Cleymans:2005xv} also fails to reproduce kaon fluctuations and over-predicts the heavy baryon fluctuations at higher center-of-mass energies.
	
	The most reasonable explanation is the sequential freeze-out of hadrons with respect to their masses. The heavier hadrons, having higher relativistic energy, hadronizes at a higher temperature compared to lighter hadrons. Although the HRG model alone is incapable of making any comment about the phase transition behavior, it effectively determines the exact thermal parameters at which the freeze-out occurs. Our results are in agreement with a previous study\cite{Adamczyk:2017iwn} where it was found that when heavier hadron yield are excluded from fitting, the $T_{ch}$ obtained is about 10 -- 20 MeV lower. 
	\section{Conclusion}
	In this paper, we find that, when using the first and second-order individual fluctuation measurements, the HRG model fits result in higher hadronization temperatures for heavier hadrons. A similar observation can be found in Ref.\cite{Adamczyk:2017iwn}, where they notice that including additional, heavier states increase the freeze-out temperature. A different picture had been proposed in Ref.\cite{Bellwied:2018tkc}, as well as Ref.\cite{Bluhm:2018aei}, where the sequential freeze-out is flavor-dependent rather than mass, leading to higher freeze-out temperature for strange hadrons such as kaon than non-strange hadrons such as proton. Although in contradiction, this work is different from either of them since it uses individual fluctuation results as well as the recently published $\Lambda$ fluctuation results\cite{Adam:2020kzk}. The model calculation was checked with Ref.\cite{Alba:2014eba}. Any decisive conclusion would require further fluctuation measurements and accurate higher-order fluctuation results.
	
	\bibliographystyle{apsrev4-2}
	\bibliography{reference} 

\begin{thebibliography}{21}%
\makeatletter
\providecommand \@ifxundefined [1]{%
 \@ifx{#1\undefined}
}%
\providecommand \@ifnum [1]{%
 \ifnum #1\expandafter \@firstoftwo
 \else \expandafter \@secondoftwo
 \fi
}%
\providecommand \@ifx [1]{%
 \ifx #1\expandafter \@firstoftwo
 \else \expandafter \@secondoftwo
 \fi
}%
\providecommand \natexlab [1]{#1}%
\providecommand \enquote  [1]{``#1''}%
\providecommand \bibnamefont  [1]{#1}%
\providecommand \bibfnamefont [1]{#1}%
\providecommand \citenamefont [1]{#1}%
\providecommand \href@noop [0]{\@secondoftwo}%
\providecommand \href [0]{\begingroup \@sanitize@url \@href}%
\providecommand \@href[1]{\@@startlink{#1}\@@href}%
\providecommand \@@href[1]{\endgroup#1\@@endlink}%
\providecommand \@sanitize@url [0]{\catcode `\\12\catcode `\$12\catcode
  `\&12\catcode `\#12\catcode `\^12\catcode `\_12\catcode `\%12\relax}%
\providecommand \@@startlink[1]{}%
\providecommand \@@endlink[0]{}%
\providecommand \url  [0]{\begingroup\@sanitize@url \@url }%
\providecommand \@url [1]{\endgroup\@href {#1}{\urlprefix }}%
\providecommand \urlprefix  [0]{URL }%
\providecommand \Eprint [0]{\href }%
\providecommand \doibase [0]{https://doi.org/}%
\providecommand \selectlanguage [0]{\@gobble}%
\providecommand \bibinfo  [0]{\@secondoftwo}%
\providecommand \bibfield  [0]{\@secondoftwo}%
\providecommand \translation [1]{[#1]}%
\providecommand \BibitemOpen [0]{}%
\providecommand \bibitemStop [0]{}%
\providecommand \bibitemNoStop [0]{.\EOS\space}%
\providecommand \EOS [0]{\spacefactor3000\relax}%
\providecommand \BibitemShut  [1]{\csname bibitem#1\endcsname}%
\let\auto@bib@innerbib\@empty
\bibitem [{\citenamefont {Braun-Munzinger}\ and\ \citenamefont
  {Wambach}(2009)}]{BraunMunzinger:2009zz}%
  \BibitemOpen
  \bibfield  {author} {\bibinfo {author} {\bibfnamefont {P.}~\bibnamefont
  {Braun-Munzinger}}\ and\ \bibinfo {author} {\bibfnamefont {J.}~\bibnamefont
  {Wambach}},\ }\href {https://doi.org/10.1103/RevModPhys.81.1031} {\bibfield
  {journal} {\bibinfo  {journal} {Rev. Mod. Phys.}\ }\textbf {\bibinfo {volume}
  {81}},\ \bibinfo {pages} {1031} (\bibinfo {year} {2009})},\ \Eprint
  {https://arxiv.org/abs/0801.4256} {arXiv:0801.4256 [hep-ph]} \BibitemShut
  {NoStop}%
\bibitem [{\citenamefont {Aoki}\ \emph {et~al.}(2006)\citenamefont {Aoki},
  \citenamefont {Endrodi}, \citenamefont {Fodor}, \citenamefont {Katz},\ and\
  \citenamefont {Szabo}}]{Aoki:2006we}%
  \BibitemOpen
  \bibfield  {author} {\bibinfo {author} {\bibfnamefont {Y.}~\bibnamefont
  {Aoki}}, \bibinfo {author} {\bibfnamefont {G.}~\bibnamefont {Endrodi}},
  \bibinfo {author} {\bibfnamefont {Z.}~\bibnamefont {Fodor}}, \bibinfo
  {author} {\bibfnamefont {S.}~\bibnamefont {Katz}},\ and\ \bibinfo {author}
  {\bibfnamefont {K.}~\bibnamefont {Szabo}},\ }\href
  {https://doi.org/10.1038/nature05120} {\bibfield  {journal} {\bibinfo
  {journal} {Nature}\ }\textbf {\bibinfo {volume} {443}},\ \bibinfo {pages}
  {675} (\bibinfo {year} {2006})},\ \Eprint
  {https://arxiv.org/abs/hep-lat/0611014} {arXiv:hep-lat/0611014} \BibitemShut
  {NoStop}%
\bibitem [{\citenamefont {Cleymans}\ \emph {et~al.}(2006)\citenamefont
  {Cleymans}, \citenamefont {Oeschler}, \citenamefont {Redlich},\ and\
  \citenamefont {Wheaton}}]{Cleymans:2005xv}%
  \BibitemOpen
  \bibfield  {author} {\bibinfo {author} {\bibfnamefont {J.}~\bibnamefont
  {Cleymans}}, \bibinfo {author} {\bibfnamefont {H.}~\bibnamefont {Oeschler}},
  \bibinfo {author} {\bibfnamefont {K.}~\bibnamefont {Redlich}},\ and\ \bibinfo
  {author} {\bibfnamefont {S.}~\bibnamefont {Wheaton}},\ }\href
  {https://doi.org/10.1103/PhysRevC.73.034905} {\bibfield  {journal} {\bibinfo
  {journal} {Phys. Rev.}\ }\textbf {\bibinfo {volume} {C73}},\ \bibinfo {pages}
  {034905} (\bibinfo {year} {2006})},\ \Eprint
  {https://arxiv.org/abs/hep-ph/0511094} {arXiv:hep-ph/0511094 [hep-ph]}
  \BibitemShut {NoStop}%
\bibitem [{\citenamefont {Braun-Munzinger}\ \emph {et~al.}(2003)\citenamefont
  {Braun-Munzinger}, \citenamefont {Redlich},\ and\ \citenamefont
  {Stachel}}]{BraunMunzinger:2003zd}%
  \BibitemOpen
  \bibfield  {author} {\bibinfo {author} {\bibfnamefont {P.}~\bibnamefont
  {Braun-Munzinger}}, \bibinfo {author} {\bibfnamefont {K.}~\bibnamefont
  {Redlich}},\ and\ \bibinfo {author} {\bibfnamefont {J.}~\bibnamefont
  {Stachel}},\ }\Eprint {https://arxiv.org/abs/nucl-th/0304013}
  {arXiv:nucl-th/0304013}  (\bibinfo {year} {2003})\BibitemShut {NoStop}%
\bibitem [{\citenamefont {Bazavov}\ \emph {et~al.}(2012)\citenamefont {Bazavov}
  \emph {et~al.}}]{Bazavov:2012jq}%
  \BibitemOpen
  \bibfield  {author} {\bibinfo {author} {\bibfnamefont {A.}~\bibnamefont
  {Bazavov}} \emph {et~al.} (\bibinfo {collaboration} {HotQCD}),\ }\href
  {https://doi.org/10.1103/PhysRevD.86.034509} {\bibfield  {journal} {\bibinfo
  {journal} {Phys. Rev.}\ }\textbf {\bibinfo {volume} {D86}},\ \bibinfo {pages}
  {034509} (\bibinfo {year} {2012})},\ \Eprint
  {https://arxiv.org/abs/1203.0784} {arXiv:1203.0784 [hep-lat]} \BibitemShut
  {NoStop}%
\bibitem [{\citenamefont {Stephanov}(2009)}]{Stephanov:2008qz}%
  \BibitemOpen
  \bibfield  {author} {\bibinfo {author} {\bibfnamefont {M.}~\bibnamefont
  {Stephanov}},\ }\href {https://doi.org/10.1103/PhysRevLett.102.032301}
  {\bibfield  {journal} {\bibinfo  {journal} {Phys. Rev. Lett.}\ }\textbf
  {\bibinfo {volume} {102}},\ \bibinfo {pages} {032301} (\bibinfo {year}
  {2009})},\ \Eprint {https://arxiv.org/abs/0809.3450} {arXiv:0809.3450
  [hep-ph]} \BibitemShut {NoStop}%
\bibitem [{\citenamefont {Bzdak}\ \emph {et~al.}(2020)\citenamefont {Bzdak},
  \citenamefont {Esumi}, \citenamefont {Koch}, \citenamefont {Liao},
  \citenamefont {Stephanov},\ and\ \citenamefont {Xu}}]{Bzdak:2019pkr}%
  \BibitemOpen
  \bibfield  {author} {\bibinfo {author} {\bibfnamefont {A.}~\bibnamefont
  {Bzdak}}, \bibinfo {author} {\bibfnamefont {S.}~\bibnamefont {Esumi}},
  \bibinfo {author} {\bibfnamefont {V.}~\bibnamefont {Koch}}, \bibinfo {author}
  {\bibfnamefont {J.}~\bibnamefont {Liao}}, \bibinfo {author} {\bibfnamefont
  {M.}~\bibnamefont {Stephanov}},\ and\ \bibinfo {author} {\bibfnamefont
  {N.}~\bibnamefont {Xu}},\ }\href
  {https://doi.org/10.1016/j.physrep.2020.01.005} {\bibfield  {journal}
  {\bibinfo  {journal} {Phys. Rept.}\ }\textbf {\bibinfo {volume} {853}},\
  \bibinfo {pages} {1} (\bibinfo {year} {2020})},\ \Eprint
  {https://arxiv.org/abs/1906.00936} {arXiv:1906.00936 [nucl-th]} \BibitemShut
  {NoStop}%
\bibitem [{\citenamefont {Adamczyk}\ \emph
  {et~al.}(2014{\natexlab{a}})\citenamefont {Adamczyk} \emph
  {et~al.}}]{Adamczyk:2013dal}%
  \BibitemOpen
  \bibfield  {author} {\bibinfo {author} {\bibfnamefont {L.}~\bibnamefont
  {Adamczyk}} \emph {et~al.} (\bibinfo {collaboration} {STAR}),\ }\href
  {https://doi.org/10.1103/PhysRevLett.112.032302} {\bibfield  {journal}
  {\bibinfo  {journal} {Phys. Rev. Lett.}\ }\textbf {\bibinfo {volume} {112}},\
  \bibinfo {pages} {032302} (\bibinfo {year} {2014}{\natexlab{a}})},\ \Eprint
  {https://arxiv.org/abs/1309.5681} {arXiv:1309.5681 [nucl-ex]} \BibitemShut
  {NoStop}%
\bibitem [{\citenamefont {Adamczyk}\ \emph {et~al.}(2018)\citenamefont
  {Adamczyk} \emph {et~al.}}]{Adamczyk:2017wsl}%
  \BibitemOpen
  \bibfield  {author} {\bibinfo {author} {\bibfnamefont {L.}~\bibnamefont
  {Adamczyk}} \emph {et~al.} (\bibinfo {collaboration} {STAR}),\ }\href
  {https://doi.org/10.1016/j.physletb.2018.07.066} {\bibfield  {journal}
  {\bibinfo  {journal} {Phys. Lett.}\ }\textbf {\bibinfo {volume} {B785}},\
  \bibinfo {pages} {551} (\bibinfo {year} {2018})},\ \Eprint
  {https://arxiv.org/abs/1709.00773} {arXiv:1709.00773 [nucl-ex]} \BibitemShut
  {NoStop}%
\bibitem [{\citenamefont {Adamczyk}\ \emph
  {et~al.}(2014{\natexlab{b}})\citenamefont {Adamczyk} \emph
  {et~al.}}]{Adamczyk:2014fia}%
  \BibitemOpen
  \bibfield  {author} {\bibinfo {author} {\bibfnamefont {L.}~\bibnamefont
  {Adamczyk}} \emph {et~al.} (\bibinfo {collaboration} {STAR}),\ }\href
  {https://doi.org/10.1103/PhysRevLett.113.092301} {\bibfield  {journal}
  {\bibinfo  {journal} {Phys. Rev. Lett.}\ }\textbf {\bibinfo {volume} {113}},\
  \bibinfo {pages} {092301} (\bibinfo {year} {2014}{\natexlab{b}})},\ \Eprint
  {https://arxiv.org/abs/1402.1558} {arXiv:1402.1558 [nucl-ex]} \BibitemShut
  {NoStop}%
\bibitem [{\citenamefont {Adam}\ \emph {et~al.}(2020)\citenamefont {Adam} \emph
  {et~al.}}]{Adam:2020kzk}%
  \BibitemOpen
  \bibfield  {author} {\bibinfo {author} {\bibfnamefont {J.}~\bibnamefont
  {Adam}} \emph {et~al.} (\bibinfo {collaboration} {STAR}),\ }\href
  {https://doi.org/10.1103/PhysRevC.102.024903} {\bibfield  {journal} {\bibinfo
   {journal} {Phys. Rev. C}\ }\textbf {\bibinfo {volume} {102}},\ \bibinfo
  {pages} {024903} (\bibinfo {year} {2020})},\ \Eprint
  {https://arxiv.org/abs/2001.06419} {arXiv:2001.06419 [nucl-ex]} \BibitemShut
  {NoStop}%
\bibitem [{\citenamefont {Bellwied}(2018)}]{Bellwied:2017uat}%
  \BibitemOpen
  \bibfield  {author} {\bibinfo {author} {\bibfnamefont {R.}~\bibnamefont
  {Bellwied}},\ }\href {https://doi.org/10.1051/epjconf/201817102006}
  {\bibfield  {journal} {\bibinfo  {journal} {EPJ Web Conf.}\ }\textbf
  {\bibinfo {volume} {171}},\ \bibinfo {pages} {02006} (\bibinfo {year}
  {2018})},\ \Eprint {https://arxiv.org/abs/1711.00514} {arXiv:1711.00514
  [nucl-ex]} \BibitemShut {NoStop}%
\bibitem [{\citenamefont {Huovinen}\ and\ \citenamefont
  {Petreczky}(2010)}]{Huovinen:2009yb}%
  \BibitemOpen
  \bibfield  {author} {\bibinfo {author} {\bibfnamefont {P.}~\bibnamefont
  {Huovinen}}\ and\ \bibinfo {author} {\bibfnamefont {P.}~\bibnamefont
  {Petreczky}},\ }\href {https://doi.org/10.1016/j.nuclphysa.2010.02.015}
  {\bibfield  {journal} {\bibinfo  {journal} {Nucl. Phys.}\ }\textbf {\bibinfo
  {volume} {A837}},\ \bibinfo {pages} {26} (\bibinfo {year} {2010})},\ \Eprint
  {https://arxiv.org/abs/0912.2541} {arXiv:0912.2541 [hep-ph]} \BibitemShut
  {NoStop}%
\bibitem [{\citenamefont {Olive}\ \emph {et~al.}(2014)\citenamefont {Olive}
  \emph {et~al.}}]{Agashe:2014kda}%
  \BibitemOpen
  \bibfield  {author} {\bibinfo {author} {\bibfnamefont {K.~A.}\ \bibnamefont
  {Olive}} \emph {et~al.} (\bibinfo {collaboration} {Particle Data Group}),\
  }\href {https://doi.org/10.1088/1674-1137/38/9/090001} {\bibfield  {journal}
  {\bibinfo  {journal} {Chin. Phys.}\ }\textbf {\bibinfo {volume} {C38}},\
  \bibinfo {pages} {090001} (\bibinfo {year} {2014})}\BibitemShut {NoStop}%
\bibitem [{\citenamefont {Fu}(2013)}]{Fu:2013gga}%
  \BibitemOpen
  \bibfield  {author} {\bibinfo {author} {\bibfnamefont {J.}~\bibnamefont
  {Fu}},\ }\href {https://doi.org/10.1016/j.physletb.2013.04.018} {\bibfield
  {journal} {\bibinfo  {journal} {Phys. Lett.}\ }\textbf {\bibinfo {volume}
  {B722}},\ \bibinfo {pages} {144} (\bibinfo {year} {2013})}\BibitemShut
  {NoStop}%
\bibitem [{\citenamefont {Nahrgang}\ \emph {et~al.}(2015)\citenamefont
  {Nahrgang}, \citenamefont {Bluhm}, \citenamefont {Alba}, \citenamefont
  {Bellwied},\ and\ \citenamefont {Ratti}}]{Nahrgang:2014fza}%
  \BibitemOpen
  \bibfield  {author} {\bibinfo {author} {\bibfnamefont {M.}~\bibnamefont
  {Nahrgang}}, \bibinfo {author} {\bibfnamefont {M.}~\bibnamefont {Bluhm}},
  \bibinfo {author} {\bibfnamefont {P.}~\bibnamefont {Alba}}, \bibinfo {author}
  {\bibfnamefont {R.}~\bibnamefont {Bellwied}},\ and\ \bibinfo {author}
  {\bibfnamefont {C.}~\bibnamefont {Ratti}},\ }\href
  {https://doi.org/10.1140/epjc/s10052-015-3775-0} {\bibfield  {journal}
  {\bibinfo  {journal} {Eur. Phys. J.}\ }\textbf {\bibinfo {volume} {C75}},\
  \bibinfo {pages} {573} (\bibinfo {year} {2015})},\ \Eprint
  {https://arxiv.org/abs/1402.1238} {arXiv:1402.1238 [hep-ph]} \BibitemShut
  {NoStop}%
\bibitem [{\citenamefont {Wheaton}\ \emph {et~al.}(2011)\citenamefont
  {Wheaton}, \citenamefont {Cleymans},\ and\ \citenamefont
  {Hauer}}]{Wheaton:2011rw}%
  \BibitemOpen
  \bibfield  {author} {\bibinfo {author} {\bibfnamefont {S.}~\bibnamefont
  {Wheaton}}, \bibinfo {author} {\bibfnamefont {J.}~\bibnamefont {Cleymans}},\
  and\ \bibinfo {author} {\bibfnamefont {M.}~\bibnamefont {Hauer}},\ }\Eprint
  {https://arxiv.org/abs/1108.4588} {arXiv:1108.4588 [hep-ph]}  (\bibinfo
  {year} {2011})\BibitemShut {NoStop}%
\bibitem [{\citenamefont {Alba}\ \emph {et~al.}(2014)\citenamefont {Alba},
  \citenamefont {Alberico}, \citenamefont {Bellwied}, \citenamefont {Bluhm},
  \citenamefont {Mantovani~Sarti}, \citenamefont {Nahrgang},\ and\
  \citenamefont {Ratti}}]{Alba:2014eba}%
  \BibitemOpen
  \bibfield  {author} {\bibinfo {author} {\bibfnamefont {P.}~\bibnamefont
  {Alba}}, \bibinfo {author} {\bibfnamefont {W.}~\bibnamefont {Alberico}},
  \bibinfo {author} {\bibfnamefont {R.}~\bibnamefont {Bellwied}}, \bibinfo
  {author} {\bibfnamefont {M.}~\bibnamefont {Bluhm}}, \bibinfo {author}
  {\bibfnamefont {V.}~\bibnamefont {Mantovani~Sarti}}, \bibinfo {author}
  {\bibfnamefont {M.}~\bibnamefont {Nahrgang}},\ and\ \bibinfo {author}
  {\bibfnamefont {C.}~\bibnamefont {Ratti}},\ }\href
  {https://doi.org/10.1016/j.physletb.2014.09.052} {\bibfield  {journal}
  {\bibinfo  {journal} {Phys. Lett. B}\ }\textbf {\bibinfo {volume} {738}},\
  \bibinfo {pages} {305} (\bibinfo {year} {2014})},\ \Eprint
  {https://arxiv.org/abs/1403.4903} {arXiv:1403.4903 [hep-ph]} \BibitemShut
  {NoStop}%
\bibitem [{\citenamefont {Bluhm}\ and\ \citenamefont
  {Nahrgang}(2019)}]{Bluhm:2018aei}%
  \BibitemOpen
  \bibfield  {author} {\bibinfo {author} {\bibfnamefont {M.}~\bibnamefont
  {Bluhm}}\ and\ \bibinfo {author} {\bibfnamefont {M.}~\bibnamefont
  {Nahrgang}},\ }\href {https://doi.org/10.1140/epjc/s10052-019-6661-3}
  {\bibfield  {journal} {\bibinfo  {journal} {Eur. Phys. J. C}\ }\textbf
  {\bibinfo {volume} {79}},\ \bibinfo {pages} {155} (\bibinfo {year} {2019})},\
  \Eprint {https://arxiv.org/abs/1806.04499} {arXiv:1806.04499 [nucl-th]}
  \BibitemShut {NoStop}%
\bibitem [{\citenamefont {Adamczyk}\ \emph {et~al.}(2017)\citenamefont
  {Adamczyk} \emph {et~al.}}]{Adamczyk:2017iwn}%
  \BibitemOpen
  \bibfield  {author} {\bibinfo {author} {\bibfnamefont {L.}~\bibnamefont
  {Adamczyk}} \emph {et~al.} (\bibinfo {collaboration} {STAR}),\ }\href
  {https://doi.org/10.1103/PhysRevC.96.044904} {\bibfield  {journal} {\bibinfo
  {journal} {Phys. Rev. C}\ }\textbf {\bibinfo {volume} {96}},\ \bibinfo
  {pages} {044904} (\bibinfo {year} {2017})},\ \Eprint
  {https://arxiv.org/abs/1701.07065} {arXiv:1701.07065 [nucl-ex]} \BibitemShut
  {NoStop}%
\bibitem [{\citenamefont {Bellwied}\ \emph {et~al.}(2019)\citenamefont
  {Bellwied}, \citenamefont {Noronha-Hostler}, \citenamefont {Parotto},
  \citenamefont {Portillo~Vazquez}, \citenamefont {Ratti},\ and\ \citenamefont
  {Stafford}}]{Bellwied:2018tkc}%
  \BibitemOpen
  \bibfield  {author} {\bibinfo {author} {\bibfnamefont {R.}~\bibnamefont
  {Bellwied}}, \bibinfo {author} {\bibfnamefont {J.}~\bibnamefont
  {Noronha-Hostler}}, \bibinfo {author} {\bibfnamefont {P.}~\bibnamefont
  {Parotto}}, \bibinfo {author} {\bibfnamefont {I.}~\bibnamefont
  {Portillo~Vazquez}}, \bibinfo {author} {\bibfnamefont {C.}~\bibnamefont
  {Ratti}},\ and\ \bibinfo {author} {\bibfnamefont {J.~M.}\ \bibnamefont
  {Stafford}},\ }\href {https://doi.org/10.1103/PhysRevC.99.034912} {\bibfield
  {journal} {\bibinfo  {journal} {Phys. Rev.}\ }\textbf {\bibinfo {volume}
  {C99}},\ \bibinfo {pages} {034912} (\bibinfo {year} {2019})},\ \Eprint
  {https://arxiv.org/abs/1805.00088} {arXiv:1805.00088 [hep-ph]} \BibitemShut
  {NoStop}%
\end{thebibliography}%
\end{document}